\documentclass{article}

\usepackage{arxiv}

\usepackage[utf8]{inputenc} 
\usepackage[T1]{fontenc}    
\usepackage{hyperref}       
\usepackage{url}            
\usepackage{booktabs}       
\usepackage{amsfonts}       
\usepackage{nicefrac}       
\usepackage{microtype}      
\usepackage{lipsum}		
\usepackage{graphicx}
\usepackage{natbib}
\usepackage{doi}
\usepackage{subcaption}
\usepackage{float}
\usepackage{cite}
\usepackage{comment}
\usepackage{amsmath,amssymb,amsfonts}
\usepackage{algorithmic}
\usepackage{graphicx}
\usepackage{textcomp}
\usepackage{multirow}
\usepackage{dblfloatfix}
\usepackage{booktabs,siunitx}
\usepackage[symbol]{footmisc}
\usepackage{tablefootnote}
\usepackage{enumitem}

\def\BibTeX{{\rm B\kern-.05em{\sc i\kern-.025em b}\kern-.08em
    T\kern-.1667em\lower.7ex\hbox{E}\kern-.125emX}}
\usepackage{array}
\newcolumntype{L}[1]{>{\raggedright\let\newline\\\arraybackslash\hspace{0pt}}m{#1}}
\newcolumntype{C}[1]{>{\centering\let\newline\\\arraybackslash\hspace{0pt}}m{#1}}
\newcolumntype{R}[1]{>{\raggedleft\let\newline\\\arraybackslash\hspace{0pt}}m{#1}}

\title{Analyzing the Effect of Data Impurity on the Detection Performances of Mental Disorders}

\author{{Rohan Kumar Gupta,  Rohit Sinha}\\
	Department of Electronics and Electrical Engineering,
Indian Institute of Technology Guwahati, Guwahati, India-781039 \\
	\texttt{(rohan\_kumar, rsinha)@iitg.ac.in} \\
}

\renewcommand{\shorttitle}{Analyzing the Effect of Data Impurity on the Detection Performances of Mental Disorders}

\hypersetup{
pdftitle={A template for the arxiv style},
pdfsubject={q-bio.NC, q-bio.QM},
pdfauthor={David S.~Hippocampus, Elias D.~Striatum},
pdfkeywords={Conventional machine learning models, Representation learning},
}

\begin{document}
\maketitle

\begin{abstract}
 The primary method for identifying mental disorders automatically has traditionally involved using binary classifiers. These classifiers are trained using behavioral data obtained from an interview setup. In this training process, data from individuals with the specific disorder under consideration are categorized as the positive class, while data from all other participants constitute the negative class. In practice, it is widely recognized that certain mental disorders share similar symptoms, causing the collected behavioral data to encompass a variety of attributes associated with multiple disorders. Consequently, attributes linked to the targeted mental disorder might also be present within the negative class.
 This data impurity may lead to sub-optimal training of the classifier for a mental disorder of interest. In this study, we investigate this hypothesis in the context of major depressive disorder (MDD) and post-traumatic stress disorder detection (PTSD). The results show that upon removal of such data impurity, MDD and PTSD detection performances are significantly improved.
\end{abstract}


\section{Introduction}
Lately, there has been a significant surge in research focused on the automated identification of mental disorders~\citep{ReviewMHD}. To create a mental disorder detection system, behavioral data is gathered from participants via interviews facilitated by either a human interviewer or a computer agent. These interviews primarily consist of a series of questions directly relevant to the specific mental disorder being targeted. Behavioral data from the participants is collected using separate modalities, such as audio~\citep{DAIC13, DAIC14, PHQ8BasedMDD, KatharinaMDDPTSD20}, video~\citep{DAIC13, DAIC14, PHQ8BasedMDD}, and physiological signals~\citep{PhysilogicalSignalsMD,ReviewPhysilohical18}. Numerous studies have reported on these modalities, focusing on detecting specific mental disorders~\citep{ReviewPhysilohical18,AVEC16, DepAudioNet}. Within these studies, the ground truth is primarily established through participants' responses to self-reported questionnaires specific to the targeted mental disorder. The criteria utilized in these self-reported questionnaires adhere to the guidelines outlined in the Diagnostic and Statistical Manual of Mental Disorders (DSM)~\citep{DSM5Book}.
This process involves identifying the minimum number of symptoms from a predefined set to diagnose a mental disorder. In the DSM, it can be observed that some mental disorders share a few symptoms. Consequently, during a clinical interview, a participant with the targeted mental disorder may exhibit behavioral data that partially resembles participants with related mental disorder(s). Among the reported studies, the mental disorder detector is typically crafted using a binary classifier, where data from individuals both with and without the specific disorder form the positive and negative classes respectively. However, the inclusion of data from participants without the targeted disorder but with related mental disorder(s) in the negative class could potentially result in elevated misclassification rates.

Major depressive disorder (MDD) is the most widely investigated mental disorder. On the other hand, post-traumatic stress disorder (PTSD) is another commonly occurring mental disorder, though relatively less studied. In the 5\textsuperscript{th} version of DSM~\citep{DSM5Book}, the sets of symptoms characterizing MDD and PTSD have four common symptoms. The authors in~\citep{PTSDsubtypeMDD21} have conducted a genetic analysis to investigate the relation between MDD and PTSD. It is found that PTSD is the subtype of MDD. Based on the correlation analysis of the responses to the self-reported questionnaires, it is reported that MDD and PTSD are highly correlated~\citep{CorrMDDPTSD}. Thus, these reported findings indicate that the characteristics of MDD and PTSD are somewhat similar. The Distress Analysis Interview Corpus Wizard-of-Oz (DAIC-WOZ)~\citep{DAIC14} is one of the most widely referred publicly available audio-video datasets. In DAIC-WOZ, each participant's data is labeled for both presence and absence of MDD and PTSD. The majority of the reported research works performed the detection of MDD without referring to the labels of PTSD~\citep{AVEC16,DepAudioNet,DialatedCNN_MDD20,FVTC_CNN_MDD20}. Upon analyzing the labels of the DAIC-WOZ dataset, we found that the data from a few non-depressed participants are labeled for PTSD. Similarly, the data from a few participants who are not labeled for PTSD but for MDD. Since MDD and PTSD are highly correlated and share some symptoms, it is possible that some acoustic and visual markers are shared between the MDD- and PTSD-diagnosed populations in the DAIC-WOZ dataset. As a result of that a binary classifier trained for MDD detection may lead to increase misclassification rates if the data from PTSD-diagnosed participants is included in the negative class. Degradation may be observed for PTSD detection performance if the data from MDD-diagnosed participants is included in the negative class.

\begin{figure}
\centering
\begin{subfigure}[b]{.5\textwidth}
  \includegraphics[width=0.75\linewidth]{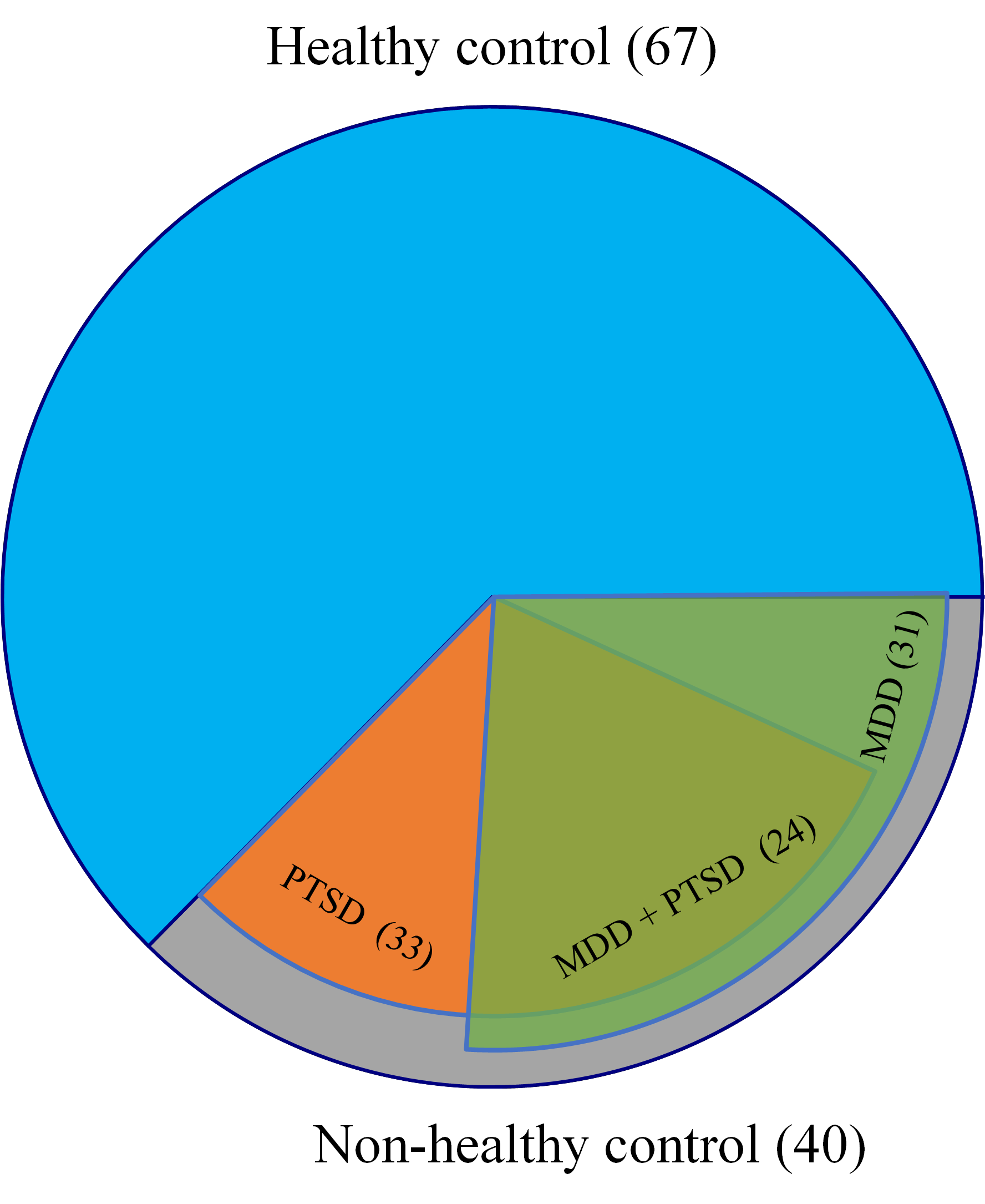}
  \caption{}
  \label{fig:sub1}
\end{subfigure}%
\begin{subfigure}[b]{.5\textwidth}
  \includegraphics[width=0.75\linewidth]{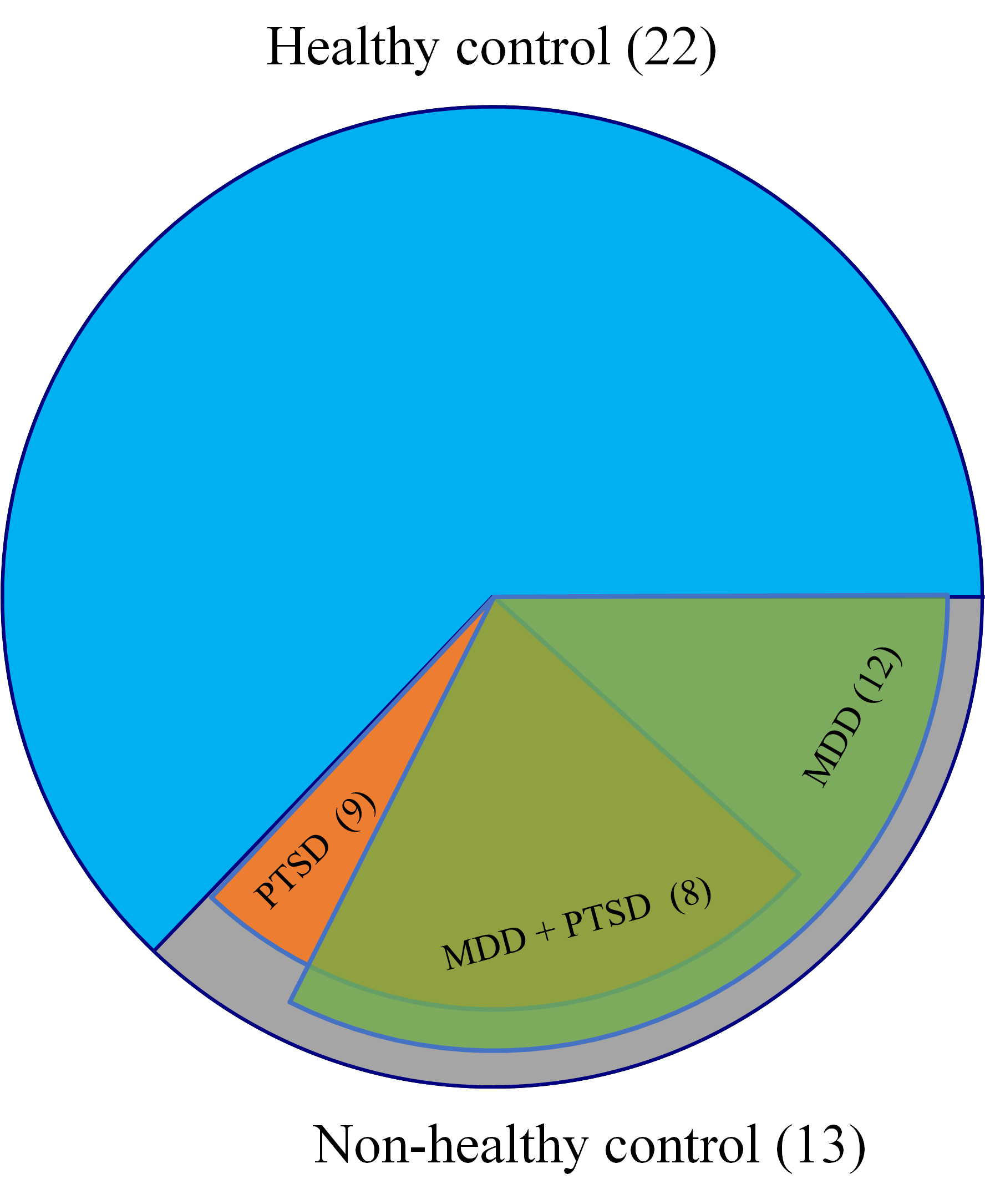}
  \caption{}
  \label{fig:sub2}
\end{subfigure}
\begin{subfigure}[b]{.5\textwidth}
  \includegraphics[width=0.75\linewidth]{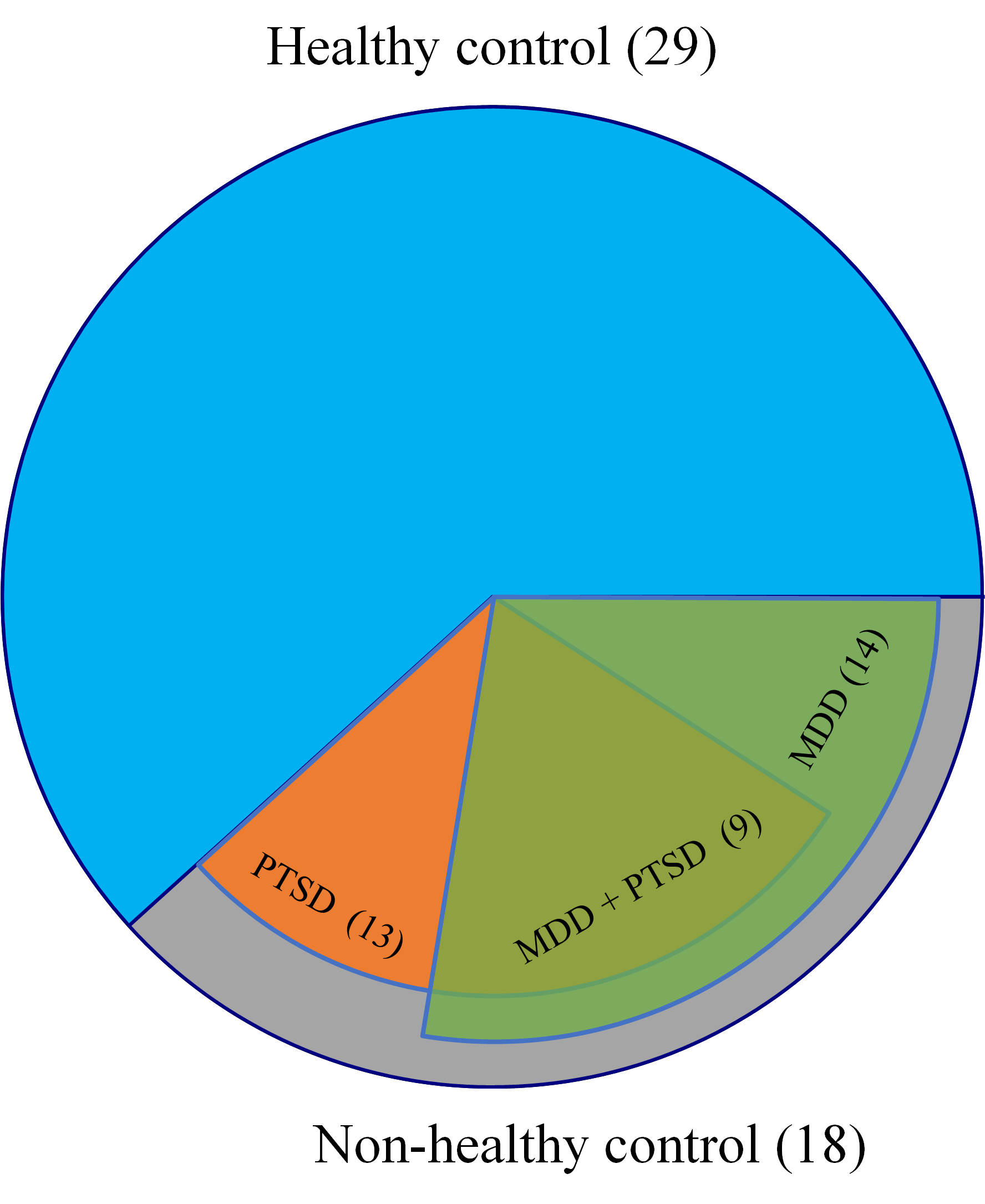}
  \caption{}
  \label{fig:sub3}
\end{subfigure}
\caption{Distribution of the participants in the (a) training, (b) development, and (c) test partitions in the DAIC-WOZ dataset. The number in the brackets indicates the population of the participants in a particular group.}
\label{fig:DAICWOZ_dis}
\end{figure}

In this study, we aim to analyze the effect of existing instances of a related mental disorder in the negative class on the binary classification performance for the detection of a mental disorder of interest. For the same, we utilized the DAIC-WOZ dataset which is labeled for MDD and PTSD. Both disorders are assigned as the primary and related mental disorders interchangeably. The rest of the paper is organized as follows. Section~\ref{Sec_exp_details} provides the experimental details. The experimental results are presented in Section~\ref{Results}. The final remarks are provided in Section~\ref{conclusions}.

\section{Experimental details}
\label{Sec_exp_details}
In the following, we provided details of the dataset, methodology, and model architectures employed in this study.

\subsection{Dataset}
This study utilized the DAIC-WOZ dataset~\citep{DAIC14}. It comprises 189 recorded interactions in audio-video format between a virtual interviewer and the participants. Labeling the audio-video data is based on participants' responses to self-reported questionnaires. The Patient Health Questionnaire (PHQ-8)\citep{PHQ8} and PTSD Checklist -- Civilian version (PCL-C)\citep{PCLC1998} questionnaires were utilized to obtain labels for the detection of MDD and PTSD, respectively. A participant is diagnosed with the disorders if their score on the questionnaires exceeds a predefined threshold which is  10 and 45 for PHQ-8 and PCL-C questionnaires, respectively.

The dataset is further divided into training, development, and test partition comprising data from 107, 35, and 47 participants, respectively. Figure~\ref{fig:sub1}, Figure~\ref{fig:sub2}, and Figure~\ref{fig:sub3} show the distribution of the participants in the training, development, and test partitions, respectively, among `healthy control' and `non-healthy control' groups. The healthy control refers to the group comprising data from participants neither diagnosed with MDD nor PTSD. In contrast, the non-healthy control group comprises data from participants diagnosed with MDD or PTSD, or both. For MDD detection, a binary classifier has been utilized that includes classes consisting of data from participants with or without MDD. Upon analyzing the partitions of the DAIC-WOZ dataset, we found that the negative class of the MDD classifier comprises the data from 9, 1, and 4 participants with PTSD in the training, development, and test partitions, respectively, along with data from the healthy control group participants. The same can also be deduced from Figure~\ref{fig:DAICWOZ_dis}. Similarly, for PTSD detection, the negative comprises data from 7, 4, and 5 participants with MDD in the training, development, and test partitions, respectively, along with data from the healthy control group participants.

\subsection{Methodology}
\label{Methodology}
This study is performed to analyze the impact on the detection performance of a mental disorder of interest due to the existence of another related mental disorder in the targeted population. The experiments are performed for MDD as being the primary disorder and PTSD as a related disorder, and vice-versa. For this study, we propose a modification that involves the removal of data of the participants with a related mental disorder from the negative class of a binary classifier trained to detect the primary mental disorder. When MDD is selected as the primary disorder, the data of the participants diagnosed with PTSD in the negative class is removed from the training partition. Similarly, when PTSD is selected as the primary disorder, the data of the participants diagnosed with MDD in the negative class is removed from the training partition. The development and test partitions are remained unaltered in all experiments. For MDD as the primary mental disorder, the resultant training partition comprises data from 98 participants. Whereas, for PTSD as the primary mental disorder, the resultant training partition comprises data from 100 participants. The baseline of these detection performances is created by training the models on the original partition of DAIC-WOZ.

\subsection{Model architectures}

\begin{figure}[t]
\begin{center}
\centerline{\includegraphics[width=8cm,height=8cm,keepaspectratio]{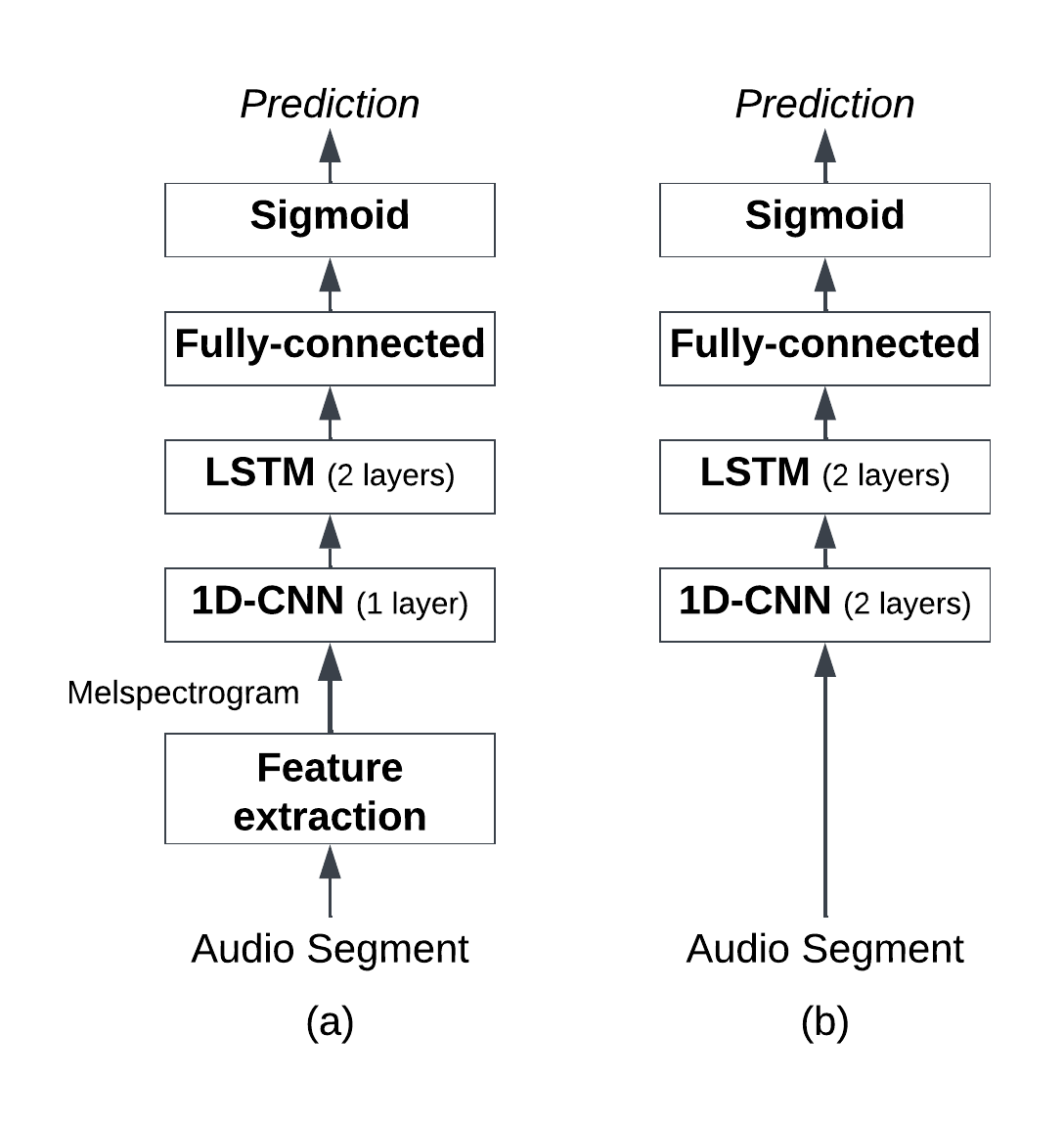}}
\caption{The schematic diagram of the model architecture of (a) `DepAudioNet', (b) `Raw Audio'.}
\label{Models}
\end{center}
\end{figure}

This study utilizes two hybrid deep learning architectures proposed in~\citep{DepAudioNet} and ~\citep{GenderBiasDepression}, and referred to as `DepAudioNet' and `Raw Audio', respectively. We used the source code of the models developed by the authors in ~\citep{GenderBiasDepression}. The schematic diagram of the models is depicted in Figure~\ref{Models}. The models are excited either by raw audio input or by low-level descriptors (LLDs) of the raw audio input. The raw audio is first pre-processed based on the methods outlined in~\citep{DepAudioNet}. The pre-processing step comprises the removal of long pauses in each audio file by utilizing the onset and offset time specified in the corresponding transcript file. The remaining portions are then concatenated together in the original sequence. The audio input is a non-overlapping segment of 61440 samples, corresponding to 3.84 sec. For the DepAudioNet model, a feature extraction step is referred to that outputs mel-filter bank features as LLDs, with 40 banks being selected. The Hanning window length used for creating LLDs is set to 64 msec, with a corresponding hop size of 32 msec. The extracted LLDs are fed to the one-dimensional convolutional neural network (1D-CNN) with kernel size 3 and stride 1 of the DepAudioNet.
In contrast, for the Raw Audio model, the audio segment is directly fed to the 1D-CNN with kernel size 1024 and stride 512. The Raw Audio model used two convolutional layers. In which, the second 1D-CNN layer has kernel size 3 and stride 1. The 1D-CNN layer is used to capture the local variation among the low-level descriptors. All other layers in both models are kept identical. The convolutional layer(s) is followed by batch normalization, rectified linear unit activation, max pooling (kernel 3, stride 3), and dropout (0.05). These layers are not shown in Figure~\ref{Models}a and Figure~\ref{Models}b for brevity. Subsequently, two unidirectional long short-term memory (LSTM) layers are employed, having 128 hidden nodes. The LSTM layers are used for capturing the long-term dependencies. The outputs of the LSTM layers are transformed to the probability of an input segment belonging to the positive class using a fully-connected layer that comprises sigmoid non-linearity. The final prediction for a participant is made based on the majority voting rule over all segments of that participant. In both models, the binary cross-entropy loss function is utilized. The models are trained using Adam optimizer.

\section{Results}
\label{Results}
In the following, we present the detection performances of the experiments mentioned in Section~\ref{Methodology}.
In this study, for each experiment, we have trained each model five times on randomly generated training subsets, and the final outcome is obtained by averaging the probabilities predicted by those. The performances of the detectors are evaluated in terms of the F1-score of positive and negative classes along with their average on the test partition of the DAIC-WOZ dataset. The hyperparameters of the models are kept identical for all experiments. All the experiments are conducted on audio data only.

Table~\ref{tableMDD} shows MDD detection performances for considered models. In which \emph{Baseline} refers to the detection performance of the models trained using original partitions. It can be observed from the table that upon the incorporation of the proposed modification in the training partition, the detection performances improved significantly compared to that of the Baseline for both considered models. For DepAudioNet, the increment in the detection performance in terms of macro-averaging F1-score (Avg.) is found to be 15.2\%. The same is 4.4\% for the raw audio model.

Table~\ref{tablePTSD} shows the PTSD detection performances for the Baseline, the proposed modification. It can be observed from the table that upon the incorporation of the proposed modification in the training partition, the detection performances improved significantly compared to that of the Baseline for both considered models. For DepAudioNet, the increment in the detection performance in terms of macro-averaging F1-score (Avg.) is found to be 10.5\%. The same is 3.0\% for the raw audio model.

\begin{table}[t]
\begin{center}
\caption{MDD detection Performances on the test partition of the DAIC-WOZ dataset in terms of F1-score. PC: positive class, NC: negative class.}
\begin{tabular}{c c c c c c c}

\hline
\hline

\multicolumn{1}{c}{\textbf{Model}} &
 \multicolumn{3}{c}{\textbf{Baseline}} & \multicolumn{3}{c}{\textbf{Proposed}} \\
\cmidrule(lr){2-4}
\cmidrule(lr){5-7}
 &\textbf{PC} &
\textbf{NC} &  \textbf{Avg.}
&\textbf{PC} &
\textbf{NC} &  \textbf{Avg.}\\
\hline
DepAudioNet & 0.235 & 0.567 & 0.401 & 0.267 & 0.656 & \bf{0.462}\\
Raw Audio & 0.160 & 0.696 & 0.428 & 0.294 & 0.600 & \textbf{0.447}\\

\hline
\hline
\end{tabular}
\label{tableMDD}
\end{center}
\end{table}

\begin{table}[t]
\begin{center}
\caption{PTSD detection Performances on the test partition of the DAIC-WOZ dataset in terms of F1-score. PC: positive class, NC: negative class. }
\begin{tabular}{c c c c c c c}

\hline
\hline

\multicolumn{1}{c}{\textbf{Model}} &
 \multicolumn{3}{c}{\textbf{Baseline}} & \multicolumn{3}{c}{\textbf{Proposed}}\\
\cmidrule(lr){2-4}
\cmidrule(lr){5-7}
& \textbf{PC} &
\textbf{NC} &  \textbf{Avg.}
&\textbf{PC} &
\textbf{NC} &  \textbf{Avg.}\\
\hline
DepAudioNet & 0.276 & 0.677 & 0.476 & 0.345 & 0.708 & \textbf{0.526}\\
Raw Audio & 0.167 & 0.714 & 0.441 & 0.222 & 0.687 & \bf{0.454}\\

\hline
\hline
\end{tabular}
\label{tablePTSD}
\end{center}
\end{table}

\section{Conclusions}
\label{conclusions}
This study investigates the effect of existing data impurity in a dataset on the detection performances of a mental disorder of interest. For the same, a publicly available audio-video dataset labeled for MDD and PTSD detection is utilized. The experimental results on audio data show that MDD detection performances on the test partition are improved substantially upon removing the presented data impurity from the training partition of the dataset for both considered models. An identical trend is noted for PTSD detection.
The future direction would involve replicating the study on other modalities as well as on different state-of-the-art models. Exploring other mental disorders would be another future direction.

\bibliographystyle{unsrtnat}
\bibliography{references} 

\end{document}